\documentclass[nofootinbib,prd,twocolumn,showpacs,showkeys,preprintnumbers]{revtex4-1}
\usepackage{hyperref,amssymb,amsmath,mathrsfs,bm,graphicx}
\usepackage{color}
\begin{document}
\title{CYLINDRICALLY SYMMETRIC RELATIVISTIC FLUIDS: A  STUDY BASED ON STRUCTURE SCALARS}
\author{L. Herrera} 
\email{laherrera@cantv.net.ve}
\author{A. Di Prisco}
\email{adiprisc@fisica.ciens.ucv.ve}
\affiliation{Departamento de F\'\i sica Te\'orica e Historia de la Ciencia,
Universidad del Pa\'{\i}s Vasco, Bilbao, Spain}
\altaffiliation{Also at U.C.V., Caracas}
\author{J. Ospino } 
\email{jhozcrae@usal.es}
\affiliation{Departamento de Matem\'atica Aplicada,  Universidad de Salamanca, Salamanca, Spain.}
\date{\today}
\begin{abstract}
Applying the $1+3$ formalism we write down the full set of equations governing the structure and the  evolution of self--gravitating cylindrically symmetric dissipative fluids with anisotropic stresses, in terms of  scalar quantities obtained from the orthogonal splitting of the Riemann tensor (structure scalars),  in the context of general relativity. These scalars which have been shown previously (in the spherically symmetric case) to be related to fundamental properties of the fluid distribution, such as: energy density, energy density inhomogeneity, local anisotropy of pressure, dissipative flux, active gravitational mass etc, are shown here to play also a very important role in the dynamics of cylindrically symmetric fluids.  It is also shown that in the static case, all possible solutions to Einstein equations may be expressed explicitly through three of these scalars. 
\end{abstract}
\date{\today}
\pacs{04.40.-b, 04.40.Nr, 05.70.Ln, 95.30.Tg}
\keywords{Relativistic fluids, cylindrically symmetric systems, dissipative fluids, causal dissipative theories.}
\maketitle
\section{INTRODUCTION}

In a recent series of papers \cite{1cil, 2cil, 3cil, 4cil} a set of scalar functions obtained from the orthogonal splitting of the Riemann tensor and referred to as structure scalars, were introduced in the discussion about the structure and evolution of spherically symmetric fluid distributions.  Such scalars (five in the spherically symmetric case) were shown to be endowed with distinct physical meaning.

In particular  they control  inhomogeneities in the energy density \cite{1cil}, and the evolution of the expansion scalar and the shear tensor   \cite{1cil, 2cil, 3cil,4cil}. Also  in the static case all possible anisotropic solutions are determined by two structure scalars \cite{1cil}.

Furthermore, the role of electric charge and cosmological constant in  structure scalars has also been recently investigated \cite{5cil}.

It is our purpose in this work to carry on a  study on cylindrically symmetric fluid distributions based on structure scalars. The motivation to undertake such an endeavour  is provided on the one hand by the conspicuous physical meaning of such scalar quantities, and on the other hand by the instrinsec interest of cylindrically symmetric systems in general relativity (see \cite{6cil, 7cil, 8cil, 9cil, 10cil, 11cil, 12cil, 13cil, 14cil, 15cil, 17cil, 16cil,ultimacil, nuevacil} and references therein).

For doing so we shall apply the $1+3$ formalism developped in \cite{21cil, 22cil, refe1, refe2, refe3}. However we shall not follow a frame formalism but a coordinate basis approach in which the orthonormal frame is only used to identify frame components of proper vectors as scalars that can have a covariant interpretation.

We shall first define the structure scalars corresponding to a general cylindrically symmetric fluid distribution, then we shall deploy the full set of  equations governing the structure and evolution of  such a system and express them  in  terms of the above mentioned scalars. A systematic though non exhaustive  study of  these equations is carried out, including the coupling of the generalized ``Euler'' equation with a transport equation.

Besides the structure scalars, we shall also introduce a set of scalars describing the shear tensor and the magnetic and electric parts of Weyl tensor.

A subset of our system of equations will be shown to exhibit the role of structure scalars on the evolution of the expansion scalar  and the shear tensor. Another subset will relate the shearfree condition, the dissipative flux and the magnetic part of the Weyl tensor, while the last four equations of our system determine the inhomogeneity factor and its evolution.

The static case shall  be considered in detail. In general it will be shown that any static solution is defined by a triplet of structure scalars.

Finally all results will be discussed in some detail in the last section, and a list of issues deserving  further attention will be presented.

\section{Collapsing anisotropic dissipative fluid cylinders: BASIC definitions AND NOTATION}\label{basic}
We shall start by  introducing basic  definitions and notation to be used throughout the text. We shall closely follow the notation of \cite{17cil} except for the fact that  we are now considering a dissipative fluid.
We assume the general time dependent diagonal non rotating
cylindrically symmetric metric
\begin{equation}
ds^2=-A^2(dt^2-dr^2)+B^2dz^2+C^2d\phi^2, \label{3}
\end{equation}

where $A$, $B$ and $C$ are functions of $t$ and $r$. To represent
cylindrical symmetry, we impose the following ranges on the
coordinates
\begin{equation}
-\infty\leq t\leq\infty, \;\; 0\leq r, \;\; -\infty<z<\infty, \;\;
0\leq\phi\leq 2\pi, \label{4}
\end{equation}
where we assume $C=0$ at $r=0$ which is a non-singular axis. We
number the coordinates $x^0=t$, $x^1=r$, $x^2=z$ and $x^3=\phi$.

Next, let us consider a collapsing cylinder filled with anisotropic and
dissipative fluid. Our study concerns either  bounded  or unbounded (cosmological) configurations, in the former case we should further assume that the fluid is bounded  by a timelike cylindrical surface
$\Sigma$.

Thus, the energy momentum tensor is given by
\begin{equation}
T_{\alpha\beta}=(\mu+P)V_{\alpha}V_{\beta}
 +P g_{\alpha\beta}+q_{\alpha} V_\beta +q_{\beta}
 V_\alpha+\Pi_{\alpha \beta},
\label{1'}
\end{equation}
where

\begin{eqnarray}
\Pi_{\alpha\beta}=\Pi_s(S_{\alpha}S_{\beta}-\frac{1}{3}h_{\alpha \beta}) +\Pi_k(K_{\alpha}K_{\beta}-\frac{1}{3}h_{\alpha \beta})
,
\label{Pi}
 \end{eqnarray}

\begin{equation}
h_{\alpha\beta}=g_{\alpha\beta}+V_{\alpha} V_{\beta},
 \label{hab}
\end{equation}

\and
\begin{equation}
q_\alpha=qL_\alpha,
\label{nuevaq}
\end{equation}
and scalars $\Pi_s$, $\Pi_k$ and $q$,  are functions of $t$ and $r$.
\noindent

Alternatively, we may write the energy momentum tensor in the form
\begin{eqnarray}
T_{\alpha\beta}&=&(\mu+P_r)V_{\alpha}V_{\beta}
 +P_rg_{\alpha\beta}+q(L_{\alpha} V_{\beta} +L_{\beta} V_{\alpha})\nonumber \\
 &&+(P_z-P_r)S_{\alpha}S_{\beta}
 +(P_{\phi}-P_r)K_{\alpha}K_{\beta},
\label{nr}
\end{eqnarray}
where
\begin{eqnarray}
(P_z-P_r)\equiv \Pi_s&&,\qquad (P_\phi-P_r)\equiv \Pi_k,  \nonumber \\&&P\equiv\frac{P_\phi+P_z+P_r}{3}.
\label{nueva2}
\end {eqnarray}
The unitary vectors $V^\alpha, L^\alpha, S^\alpha, K^\alpha$ form a canonical  orthonormal tetrad. $V^\alpha$ is a hypersurface orthogonal $4-$velocity vector, ${S^\alpha, K^\alpha}$ are tangent to the orbits of the $2-$ dimensional group that defines cylindrical symmetry and $L^\alpha$ is orthogonal to these orbits and to $V^\alpha$.
With the above definitions it is clear that $\mu$ is the energy
density (the eigenvalue of $T_{\alpha\beta}$ for eigenvector $V^\alpha$), $q_\alpha$ is the radial heat flux, whereas  $P$ is the isotropic pressure.

We choose the fluid to be comoving in this coordinate system,
hence 
\begin{equation}
V_{\alpha}=-A\delta^0_{\alpha}, \;\;L_\alpha=A\delta ^1_\alpha,
\;\; S_{\alpha}=B\delta^2_{\alpha}, \;\;
K_{\alpha}=C\delta^3_{\alpha}. \label{5}
\end{equation}

Now, in order to avoid misunderstandings, the following remark is in order: It is well known that the choice of the vector $V^\alpha$ is not unique, we could for example choose a ``tilted'' congruence and therefore  the splitting  of the energy--momentum tensor would be different of course. In this sense our study is related  to the congruence of observers which are at rest at each point with respect to the corresponding fluid element, i.e. a congruence for which the four velocity of the fluid is $V^\alpha $  given by (\ref{5}).

With the notation above we can write nonvanishing components of the Einstein equations (see Appendix I).

\subsection{Kinematical Variables}
Since we are considering nonrotating fluid distributions, there are only three kinematical variables: the expansion
$\Theta$, the four acceleration $a^\alpha$ and the  shear $\sigma_{\alpha\beta}$  which as usual are defined by:
\begin{equation}
\Theta={V^{\alpha}}_{;\alpha}, \label{6} 
\end{equation}
\begin{equation}
a_\alpha =V_{\alpha;\beta}V^\beta . \label{7ciln}
\end{equation}
\begin{equation}
\sigma_{\alpha\beta}=V_{(\alpha;\beta)}
+a_{(\alpha}V_{\beta)}-\frac{1}{3}
\Theta h_{\alpha\beta}. \label{7}
\end{equation}

Using (\ref{3}),  (\ref{6})  and (\ref{7}) we obtain for the expansion,
\begin{equation}
\Theta=\frac{1}{A}\left(\frac{\dot{A}}{A}+\frac{\dot{B}}{B}
          +\frac{\dot{C}}{C}\right),
\label{8}
\end{equation}
and for the non zero components of the shear,
\begin{eqnarray}
\sigma_{11}&=&\frac{A}{3}\left(2\frac{\dot{A}}{A}-\frac{\dot{B}}{B}
               -\frac{\dot{C}}{C}\right),
\label{9} \\
\sigma_{22}&=&\frac{B^2}{3A}\left(2\frac{\dot{B}}{B}-\frac{\dot{A}}{A}
               -\frac{\dot{C}}{C}\right),
\label{10} \\
\sigma_{33}&=&\frac{C^2}{3A}\left(2\frac{\dot{C}}{C}-\frac{\dot{A}}{A}
                -\frac{\dot{B}}{B}\right),
\label{11}
\end{eqnarray}
where overdots and primes stand for partial differentiation with respect  to $t$ and $r$ respectively.

We can also express the shear tensor as
\begin{equation}
\sigma_{\alpha \beta} = \sigma_s \left(S_\alpha S_\beta -\frac{1}{3} h_{\alpha \beta}\right)
+ \sigma_k \left(K_\alpha K_\beta -\frac{1}{3} h_{\alpha \beta}\right),
\label{sh}
\end{equation}
with
\begin{equation}
\sigma_s=\frac{1}{A}\left(\frac{\dot B}{B}-\frac{\dot A}{A}\right),
\label{ss}
\end{equation}
\begin{equation}
\sigma_k=\frac{1}{A}\left(\frac{\dot C}{C}-\frac{\dot A}{A}\right),
\label{sk}
\end{equation}
and
\begin{equation}
\sigma^{\alpha \beta} \sigma_{\alpha \beta}= \frac{2}{3} \left(\sigma_s^2 - \sigma_s \sigma_k + \sigma_k^2\right).
\label{sig2}
\end{equation}
Observe that unlike the spherically symmetric case, the shear  depends now on two nonvanishing independent scalars.
\noindent Finally, for the four acceleration $a_\alpha$  we obtain 
\begin{equation}
a_\alpha=aL_{\alpha} \quad {\rm with} \quad
a=\frac{A^\prime}{A^2}.
\label{a}\end{equation}

Before ending this section it should be recalled that in the case of bounded configurations, junction conditions (Darmois) should be satisfied on the boundary surface in order to exclude the presence of thin shells. Such conditions  shall not be deployed here since we shall not use them but the reader may found a detailed discussion on this point in \cite{17cil}, \cite{16cil}.

\section{The orthogonal splitting of Riemann  Tensor and structure scalars}
In order to introduce the structure scalars corresponding to our problem, let us first recall that 
the Riemann tensor may be expressed trough the Weyl tensor
$C^\rho_{\,\,\alpha \beta \mu}$, the Ricci tensor $R_{\alpha
\beta}$, and the scalar curvature $R$.

The electric ($E_{\alpha\beta}$) and  magnetic ($H_{\alpha\beta}$)  parts of the Weyl tensor $C_{\alpha \beta
\gamma\delta}$, are defined as usual by
\begin{eqnarray}
E_{\alpha \beta}&=&C_{\alpha\nu\beta\delta}V^\nu V^\delta,\nonumber\\
H_{\alpha\beta}&=&\frac{1}{2}\eta_{\alpha \nu \epsilon \rho}C^{\quad
\epsilon\rho}_{\beta \delta}V^\nu V^\delta\,,\label{EH}
\end{eqnarray}
\noindent where $\eta_{\alpha \nu \epsilon \rho}$ is the Levi--Civita tensor and $\epsilon _{\alpha \beta \rho}=\eta_{\nu \alpha \beta \rho}V^\nu$.

Calculating (\ref{EH}) for the metric (\ref{3}) we
obtain
\begin{equation}
E_{\alpha\beta}=E_s (S_\alpha S_\beta-\frac{1}{3}h_{\alpha \beta})
+E_k(K_\alpha K_\beta-\frac{1}{3}h_{\alpha \beta}), \label{E'}
\end{equation}
\noindent with
\begin{eqnarray}
E_s=\frac{1}{A^2B^2}C_{0202}-\frac{1}{A^4}C_{0101}, \nonumber \\
E_k=\frac{1}{A^2C^2}C_{0303}-\frac{1}{A^4}C_{0101},\label{EE}
\end{eqnarray}
and
\begin{equation}
H_{\alpha\beta}=H(S_\alpha K_\beta+S_\beta K_\alpha)\label{H'},
\end{equation}
with
\begin{equation}
H=-\frac{C_{0313}}{A^2C^2}\label{HH},
\end{equation}
where the explicit expressions for the Weyl tensor components (computed for the orthonormal ``canonical tetrdad") may be found in the Appendix II. 

It is worth observing that now, unlike the spherically symetric case, the magnetic part of the Weyl tensor does not vanish (in general) and the electric part depends on two independent scalar functions.

Now, the orthogonal splitting of the Riemann tensor is  carried out by means of three tensors $Y_{\alpha\beta}$, $X_{\alpha\beta}$ and $Z_{\alpha\beta}$ defined as (see \cite{1cil, 18cil, 19cil} for details)

\begin{equation}
Y_{\alpha \beta}=R_{\alpha \nu \beta \delta}V^\nu V^\delta,
\label{Y}
\end{equation}
\begin{equation}
X_{\alpha \beta}=\frac{1}{2}\eta_{\alpha\nu}^{\quad \epsilon
\rho}R^\star_{\epsilon \rho \beta \delta}V^\nu
V^\delta,\label{X}
\end{equation}
and 
\begin{equation}
Z_{\alpha\beta}=\frac{1}{2}\epsilon_{\alpha \epsilon \rho}R^{\quad
\epsilon\rho}_{ \delta \beta} V^\delta,\label{Z}
\end{equation}
 where $R^\star _{\alpha \beta \nu
\delta}=\frac{1}{2}\eta_{\epsilon\rho\nu\delta}R_{\alpha
\beta}^{\quad \epsilon \rho}$.

 Using  the decomposition of the Riemann tensor in terms of the matter variables and the electric and magnetic parts of Weyl tensor (see\cite{1cil}, \cite{newcil})  and (\ref{E'})  (\ref{Y}), we obtain:
\begin{eqnarray}
Y_{\alpha \beta}=\frac{1}{3}Y_T h_{\alpha \beta} &+& Y_s (S_\alpha
S_\beta-\frac{1}{3}h_{\alpha \beta})\nonumber \\
&+&Y_k (K_\alpha
K_\beta-\frac{1}{3}h_{\alpha \beta}), \label{yf}
\end{eqnarray}
with
\begin{eqnarray}
Y_T=\frac{\kappa}{2}(\mu+P_z+P_\phi+P_r), \label{ortc1}\\
Y_s=E_s-\frac{\kappa}{2}(P_z-P_r), \label{ortc2}\\
Y_k=E_k-\frac{\kappa}{2}(P_\phi-P_r).
\label{YY}
\end{eqnarray}

In a similar way the tensor $X_{\alpha \beta}$ can be written as:
\begin{eqnarray}
X_{\alpha \beta}=\frac{1}{3}X_Th_{\alpha \beta} &+& X_s (S_\alpha
S_\beta-\frac{1}{3}h_{\alpha \beta})\nonumber \\
&+&X_k( K_\alpha
K_\beta-\frac{1}{3}h_{\alpha \beta}), \label{x1}
\end{eqnarray}
with
\begin{eqnarray}
X_T=\kappa\mu, \label{ortc3}\\
X_s=-E_s-\frac{\kappa}{2}(P_z-P_r),\label{ortc4}\\
X_k=-E_k-\frac{\kappa}{2}(P_\phi-P_r). \label{XX}
\end{eqnarray}

Finally, from (\ref{EH}), (\ref{H'}) and (\ref{Z}) we
obtain
\begin{equation}
Z_{\alpha\beta}=H_{\alpha\beta}+\frac{1}{2}\kappa q^\rho
\epsilon_{\alpha\beta \rho}. \label{Z'}
\end{equation}

From (\ref{Z'}) two  scalar functions may be defined as follows:
\begin{equation}
Z_H=2H=(S^\alpha K^\beta+S^\beta K^\alpha)Z_{\alpha
\beta},\label{o}
\end{equation}
\begin{equation}
Z_q=\kappa q=(S^\beta K^\alpha-S^\alpha K^\beta)Z_{\alpha
\beta}.\label{p}
\end{equation}

Thus  the full set of structure scalars are now defined by the eight quantities:
 $Y_T$, $Y_s$, $Y_k$, $X_T$, $X_s$, $X_k$, $Z_H$, $Z_q$. The corresponding expressions of these scalars 
in terms of the metric functions are given in the Appendix III.

Before ending this section it would be useful to introduce a relevant quantity defined in terms of  tensors $Y_{\alpha \beta}, X_{\alpha \beta}, Z_{\alpha \beta}$. This is the super--Poynting vector defined by
\begin{equation}
P_\alpha = \epsilon_{\alpha \beta \gamma}\left(Y^\gamma_\delta Z^{\beta \delta} - X^\gamma_\delta Z^{\delta\beta}\right),
\label{SPdef}
\end{equation}
in our case the above expression becomes
\begin{equation}
 P_\alpha=2H(E_k-E_s)S^\beta K^\gamma\epsilon_{\alpha \beta \gamma}
 +\frac{\kappa ^2}{2} (\mu +P_r)q_\alpha. \label{SP}
\end{equation}

This four--vector describes the flux of superenergy and, as it is apparent from (\ref{SP}), it embodies two contributions: one from the dissipative flux and the other from a term which represents gravitational radiation \cite{gr1, gr2}, this last term being proportional to the magnetic part of the Weyl tensor.

\section{Basic Equations}

In this section we shall deploy the relevant equations for describing a dissipative self--gravitating locally anisotropic, cylindrically symetric  fluid. In spite of the fact that not all these equations are
independent (for example the field equations and the conservation equations (Bianchi identities)) we shall present them all, since depending on the problem under consideration, it may
be more advantageous using one subset instead of the other. As mentioned in the Introduction these equation are obtained applying the 1+3 formalism to cylindrical symmetry   \cite{21cil, 22cil,refe1, refe2, refe3} (for the specific case of spherical symmetry see \cite{20cil}).

\subsection{Conservation Laws}

The conservation law $T^{\alpha \beta}_{;\beta}=0$ leads to the following couple of equations
\begin{widetext}
\begin{eqnarray}
D_t \mu &+&\Theta \left [ \mu+\frac{1}{3}(P_r+P_z+P_\phi)\right ]+\nabla q+
q\left
[2a+\frac{1}{A}\left(\frac{B^\prime}{B}+\frac{C^\prime}{C}\right)\right
]\nonumber \\
&+&\frac{1}{3}(P_z-P_r)(2\sigma
_s-\sigma_k)+\frac{1}{3}(P_\phi-P_r)(2\sigma_k-\sigma _s)=0,
\label{e1}
\end{eqnarray}
\end{widetext}
\begin{widetext}
\begin{eqnarray}
\nabla  P_r +D_t q-\frac{1}{A}\left[(P_z-P_r)\frac{B^\prime}{B}+(P_\phi-P_r)\frac{C^\prime}{C}\right]+(\mu+P_r)a-\frac{1}{3}(\sigma_s +\sigma _k-4\Theta) q=0, \label{e2}
\end{eqnarray}
\end{widetext}

\noindent where $D_t f=f_{,\alpha}V^\alpha$ and
$\nabla f=f_{,\alpha}L^\alpha$.

\subsection{Ricci Identities}
From the  Ricci identities for the vector $V_\alpha$
the following evolution equations for the expansion (the Raychaudhuri equation) and the shear tensor, as well as some constraint equations  are obtained:
\begin{widetext}
\begin{eqnarray}
 D_t \Theta-\nabla a-a^2-a\frac{1}{A}\left(\frac{B^\prime}{B}+\frac{C^\prime}{C}\right)+\frac{1}{3}\Theta
 ^2 +\frac{2}{3}(\sigma_s ^2-\sigma_s \sigma_k+\sigma_k ^2)
=-Y_T,\label{e3}
\end{eqnarray}
\end{widetext}
\begin{widetext}
\begin{eqnarray}
a\frac{1}{A}\left(2\frac{B^\prime}{B}-\frac{C^\prime}{C}\right)-a^2-\nabla a
-D_t(2\sigma _s-\sigma_k)+\frac{1}{3}(\sigma_k ^2-2\sigma
_s^2+2\sigma_s\sigma _k)-\frac{2}{3}\Theta
(2\sigma_s-\sigma_k)\nonumber \\=2Y_s-Y_k,\label{es5}
\end{eqnarray}
\end{widetext}
\begin{widetext}
\begin{eqnarray}
2 \nabla a
+2a^2-a\frac{1}{A}\left(\frac{B^\prime}{B}+\frac{C^\prime}{C}\right)
+D_t(\sigma_s+\sigma_k)+ \frac{1}{3}(\sigma_s^2+\sigma_k
^2-4\sigma_s
\sigma_k)\nonumber\\+\frac{2}{3}\Theta(\sigma_s+\sigma_k)=-(Y_s+Y_k),\label{es4}
\end{eqnarray}
\end{widetext}
\begin{equation}
\frac{1}{3}\nabla \left (2\Theta+\sigma _s+\sigma _k\right
)+\sigma _s\frac{B^\prime}{AB}+\sigma
_k\frac{C^\prime}{AC}=Z_q, \label{e6}
\end{equation}
\\
\begin{equation}
Z_H=\nabla (\sigma_k-\sigma_s)+\left(\sigma_k
\frac{C^\prime}{AC}-\sigma_s \frac{B^\prime}{AB}\right),\label{e7}
\end{equation}
\subsection{Differential equations for the Weyl tensor derived from  Bianchi Identities}

\noindent The Bianchi identities, using the Einstein equations,
lead to the following equations:

\begin{widetext}
\begin{eqnarray}
-\nabla (Y_s+Y_k-X_s-X_k) - 3(Y_s-X_s)\frac{B^\prime}{AB}- 3(Y_k-X_k)\frac{C^\prime}{AC}-6H(\sigma_s-\sigma_k)\nonumber\\
=\kappa \nabla (2\mu+P_r+P_z+P_\phi) +3\kappa (\mu+P_r) a +2\kappa q (\Theta -\sigma_s-\sigma_k)+3\kappa D_t q\nonumber\\
\label{e8}
\end{eqnarray}
\end{widetext}
\begin{widetext}
\begin{eqnarray}
\nabla (2Y_s-Y_k-2X_s+X_k) + 3(Y_s-X_s)\frac{B^\prime}{AB}+3a(Y_s-Y_k-X_s+X_k)+6H(\Theta-\sigma_k)\nonumber\\+6D_t H
=-\kappa \nabla (\mu-P_r-P_z+2P_\phi) -3\kappa (P_\phi-P_r) \frac{C^\prime}{AC} +\kappa q (\Theta -\sigma_s+2\sigma_k),\nonumber\\
\label{e12}
\end{eqnarray}
\end{widetext}
\begin{widetext}
\begin{eqnarray}
\sigma_s(-Y_s+2Y_k+X_s-2X_k)+\sigma_k(2Y_s-Y_k-2X_s+X_k)+\Theta(Y_s+Y_k-X_s-X_k)\nonumber\\
+D_t(Y_s+Y_k-X_s-X_k)-6H\left(\frac{B^\prime}{AB}-\frac{C^\prime}{AC}\right)\nonumber\\=
\kappa(\mu+P_r)(\Theta-\sigma_s-\sigma_k)+
\kappa D_t(\mu+2P_r-P_z-P_\phi)+3\kappa \nabla q+6\kappa q a,\nonumber\\
\label {e9}
\end{eqnarray}
\end{widetext}
\begin{widetext}
\begin{eqnarray}
\sigma_s(2Y_s-Y_k-2X_s+X_k)-\sigma_k(Y_s+Y_k-X_s-X_k)-\Theta(2Y_s-Y_k-2X_s+X_k)\nonumber\\
-D_t(2Y_s-Y_k-2X_s+X_k)-6H \frac{C^\prime}{AC}-6\nabla H-12 H a\nonumber\\=
\kappa(\mu+P_z)(\Theta+2\sigma_s-\sigma_k)+
\kappa D_t(\mu-P_r+2P_z-P_\phi)+3\kappa q \frac{B^\prime}{AB}.\nonumber\\
\label{e10}
\end{eqnarray}
\end{widetext}

We shall next proceed to analyze different problems using different subsets of the equations above.

\section{ dynamical equation,  transport equation and  thermoinertial efect}
We shall now elaborate on (\ref{e2}) (the generalized ``Euler'' equation) as follows.

By analogy with the spherically symmetric case let us first define the ``velocity'' 
\begin{equation}
U=\frac{\dot C}{A}=D_t C,
\label{U}
\end{equation}
then using (\ref{15}) we get
\begin{eqnarray}
D_t U &=&  a  \frac{C^\prime}{A}-\kappa P_r C \nonumber\\&-&
\frac{C}{A^2}\left(\frac{\ddot B}{B} - \frac{\dot A \dot B}{A B} + \frac{\dot B \dot C}{B C} - 
\frac{B^\prime C^\prime}{B C}-\frac{A^\prime B^\prime}{A B}\right),
\label{Uast}
\end{eqnarray}
that can be also written down as (see Appendix I)
\begin{eqnarray}
D_t U &=&  a \frac{C^\prime}{A} - \kappa P_r C \nonumber\\
&+& \frac{C}{B^2} \left(\frac{R_{0202}}{A^2} - \frac{R_{2323}}{C^2}\right).
\label{UastR}
\end{eqnarray}
Solving the above equation for the $a$ term and feeding this back into (\ref{e2})  we obtain
\begin{widetext}
\begin{eqnarray}
(\mu + P_r) D_t U = 
-(\mu + P_r ) \left[\kappa P_r C -  \frac{C}{B^2} \left(\frac{R_{0202}}{A^2} - \frac{R_{2323}}{C^2}\right) \right]\nonumber \\
+  \frac{C^\prime}{A} \left[- \nabla P_r+ (P_z-P_r)\frac{B^\prime}{A B}+(P_\phi-P_r)\frac{C^\prime}{A C} \right]
+  \frac{C^\prime}{A} \left[-D_t q + \frac{1}{3}(\sigma_s +\sigma _k-4\Theta) q \right],
\label{Fma}
\end{eqnarray}
\end{widetext}

The equation above has the ``Newtonian'' form
 \begin{equation}Force= Mass \; density \times Acceleration.\nonumber\label{Newton}\end{equation}

Indeed, the term on the left is the inertial mass (density) multiplied by the proper time derivative of  the ``velocity'' $U$. On the right hand we have three different terms: the first one is just the ``gravitational force'' term (see below), the second one includes hydrodynamic force terms (presure gradient plus the anisotropic contributions), finally the last term represent the contribution from dissipative processes.

Now, before proceeding further, let us identify  the first term on the right of (\ref{Fma}) as the ``gravitational force''.  For doing that, let us consider the static vacuum case (Levi-Civita).
From Einstein equations (\ref{13})--(\ref{17}) it follows that the metric (\ref{3}) in this case has components:
\begin{equation}
B=\frac{\alpha}{r^\beta},
\label{1m}
\end{equation}
\begin{equation}
C=\frac{r^{(\beta+1)}}{\alpha},
\label{2m}
\end{equation}
\begin{equation}
A=\frac{r^{\beta(\beta+1)}}{\alpha},
\label{3m}
\end{equation}
with $\alpha, \beta$ constant.

Then, the term 
\begin{equation}
-\frac{C}{B^2}\left(\frac{R_{0202}}{A^2}-\frac{R_{2323}}{C^2}\right),
\label{5m}
\end{equation}
becomes
\begin{equation}
\frac{C}{A^2}\frac{\beta}{r^2}(\beta+1)^2.
\label{6m}
\end{equation}

Now in the weak field  limit  the gravitational potential of an infinite line with mass per unit of length $\sigma$ is
\begin{equation}
\Phi=2 \sigma \ln {r}+ {\rm constant}.
\label{7m}
\end{equation}

Therefore  in that limit  we have
\begin{equation}
\beta^2+\beta=2 \sigma,
\label{8m}
\end{equation}
implying
\begin{equation}
\beta=\frac{-1+\sqrt{1+8\sigma}}{2},
\label{9m}
\end{equation}
since in the weak field limit $\sigma<<1$ then 
\begin{equation}
\beta\approx 2 \sigma.
\label{10m}
\end{equation}

Thus (\ref{6m}) becomes in the weak field limit
\begin{equation}
\frac{C}{A^2}\frac{2\sigma}{r^2},
\label{11m}
\end{equation}
but in that limit  $C\approx r$, $A\approx 1$  and the term above is just  the gravitational force exerted by the infinite line.

\subsection{Thermoinertial effect}

Let us now get back to our equation (\ref{Fma}). In order to obtain an expression for the  $D_t q$ term we have  to  resort to a transport equation.
We shall need a transport equation derived from  a causal  dissipative theory ( e.g. the
M\"{u}ller-Israel-Stewart second
order phenomenological theory for dissipative fluids \cite{Muller67, IsSt76, I, II}).

Indeed, as it is already  well known the Maxwell-Fourier law for
radiation flux leads to a parabolic equation (diffusion equation)
which predicts propagation of perturbations with infinite speed
(see \cite{6D}-\cite{8'} and references therein). This simple fact
is at the origin of the pathologies \cite{9H} found in the
approaches of Eckart \cite{10E} and Landau \cite{11L} for
relativistic dissipative processes. To overcome such difficulties,
various relativistic
theories with non-vanishing relaxation times have been proposed in
the past \cite{Muller67,IsSt76, I, II, 14Di,15d}. Although the final word on this issue has not yet been said, the important point is that
all these theories provide a heat transport equation which is not
of Maxwell-Fourier type but of Cattaneo type \cite{18D}, leading
thereby to a hyperbolic equation for the propagation of thermal
perturbations (see \cite{cil23} for a recent discussion on this issue).

A key quantity in these theories is the relaxation time $\tau$ of the
corresponding  dissipative process. This positive--definite quantity has a
distinct physical meaning, namely the time taken by the system to return
spontaneously to the steady state (whether of thermodynamic equilibrium or
not) after it has been suddenly removed from it. 
Therefore, when studying transient regimes, i.e., the evolution from a 
steady--state situation to a new one,  $\tau$ cannot be neglected. 

Sometimes in the past it has been argued that dissipative processes with relaxation times
comparable  to the characteristic time of the system are out of the
hydrodynamic regime.  However, the concept of hydrodynamic regime
involves the ratio between the  mean free path of fluid particles  and
the characteristic length of the system. 
Therefore that argument  can be valid only if the particles making up the
fluid are the same ones that transport the heat. However, this is 
never the case. Specifically, for a neutron star, $\tau$ is of the
order of the scattering time between electrons (which carry the
heat) but this fact is not an obstacle (no matter how large the
mean free path of these electrons may be) to consider the neutron
star as formed by a Fermi fluid of degenerate neutrons. The same
is true for the second sound in superfluid Helium and solids, and
for almost any ordinary fluid. In brief, the hydrodynamic regime
refers to fluid particles that not necessarily (and as a matter of fact,
almost never) transport the heat. Therefore large relaxation times (large
mean free paths of particles involved in heat transport) does not imply a
departure from the hydrodynamic regime (this fact has been stressed before
\cite{Santos}, but it is usually overlooked).

Thus, the transport equation reads
\begin{eqnarray}
\tau h^{\alpha \beta} V^\gamma q_{\beta;\gamma} + q^\alpha \nonumber\\
= - K h^{\alpha \beta} \left(T_{,\beta} + T a\right) - \frac{1}{2}K T^2 \left(\frac{\tau V^\beta}{K T^2}\right)_{;\beta} q^\alpha,
\label{trans}
\end{eqnarray}
where  $K$ and $T$ denote thermal conductivity and temperature respectively, and whose only component in our case is 
\begin{eqnarray}
D_t\tau q + q =\nonumber\\
 -K (\nabla T + T a) - \frac{1}{2}K T^2 q D_t \left(\frac{\tau}{K T^2}\right) - \frac{1}{2} \tau q \Theta.
\label{qast}
\end{eqnarray}

Then feeding back  (\ref{qast}) into  (\ref{Fma}),  we obtain
\begin{widetext}
Then feeding back  (\ref{qast}) into  (\ref{Fma}),  we obtain
\begin{eqnarray}
(\mu + P_r)\left[1-\frac{K T}{\tau(\mu + P_r)}\right] D_tU = 
-(\mu + P_r )\left[1-\frac{K T}{\tau(\mu + P_r)}\right] \left(\frac{\kappa}{2} P_r C^3 +m\right) \frac{1}{C^2}\nonumber \\
+  \frac{C^\prime}{A} \left[- \nabla P_r+ (P_z-P_r)\frac{B^\prime}{A B}+(P_\phi-P_r)\frac{C^\prime}{A C} \right]
+  \frac{C^\prime}{A} \left[\frac{1}{2}\frac{K T^2}{\tau} q D_t\left(\frac{\tau}{K T^2}\right) + \frac{q}{\tau} + \frac{ K\nabla T}{\tau} + \frac{1}{3}(\sigma_s +\sigma _k) q - \frac{5}{6}q \Theta\right],
\label{Fmaq}
\end{eqnarray}
\end{widetext}
where the function $m$ is defined  by 
\begin{equation}
m=-\frac{\kappa P_r C^3}{2} -  \frac{C^3}{B^2} \left(\frac{R_{0202}}{A^2} - \frac{R_{2323}}{C^2}\right).
\label{mn}
\end{equation}
Equation (\ref{Fmaq}) indicates how inertial thermal effects reduce the  effective inertial mass. This effect  and its consequences, first reported in \cite{cild1}, has been extensively discussed in the past (see  \cite{cild2, cild3, cild4, cild5, 81, cild6, 82, 83} and references therein).

\section{evolution of the expansion scalar and the shear: the geodesic case}
Let us now turn to equations (\ref{e3}), (\ref{es5}) and (\ref{es4}). The former is just the evolution equation for the expansion scalar (Raychaudhuri equation) for the cylindrically symmetric case whereas the latter describes the evolution of the shear. For simplicity we shall restrict ourselves to the geodesic case ($a=0$).

First of all observe that the evolution of the expansion scalar is fully controlled by the scalar $Y_T$ (as in the spherically symmetric case). For the shear however we have two equations (for the two independent components of the shear tensor). Introducing the variables:
\begin{equation}
\sigma_I \equiv \sigma_s + \sigma_k \qquad;\qquad \sigma_{II} \equiv 2\sigma_s - \sigma_k,
\label{sIyII}
\end{equation}
\begin{equation}
Y_I \equiv Y_s + Y_k \qquad;\qquad Y_{II} \equiv 2 Y_s - Y_k,
\label{YIyII}
\end{equation}
(\ref{es5}) and (\ref{es4}) become (in the geodesic case)
\begin{equation}
D_t\sigma_I=-Y_I-
\frac{1}{9}(2\sigma_{II}^2-\sigma^2_I-2\sigma_I\sigma_{II})-\frac{2}{3}\Theta
\sigma_I\label{59}
\end{equation}
\begin{equation}
D_t\sigma_{II}=-Y_{II}
+\frac{1}{9}(2\sigma^2_I-2\sigma_I\sigma_{II}-\sigma_{II}^2)-\frac{2}{3}\Theta\sigma_{II}.\label{(60)}
\end{equation}

The two equations above describe the evolution of the shear tensor, which is fully controlled by the scalars $Y_I$ and $Y_{II}$ ($Y_s, Y_k$). If we assume that the fluid is initially ($t=0$) shearfree and scalars $Y_I$ and $Y_{II}$ vanish for a timelike  interval ($t_1> t\geq 0$), then it follows at once from (\ref{59}) and (\ref{(60)}) that the fluid  remains shearfree in that interval. However, even small deviations from the vanishing condition of the above mentioned scalars would produce deviations from the shearfree condition. In other words, all the information about the stability of the shearfree condition is encoded in $Y_I$ and $Y_{II}$ (for a discussion on this problem in the spherically symmetric case see \cite{3cil}).

\section{The link between  Shearfree Condition, dissipative flux and the magnetic part of the Weyl tensor}
We shall now  try to extract some of the information contained in (\ref{e6}) and (\ref{e7}). 

First of all observe that (\ref{e6}) implies that under the shearfree condition the inhomogeneity of 
the expansion scalar $\nabla \Theta$ is controlled by $Z_q$. Also from (\ref{e7}) it follows 
at once that the shearfree condition implies that the magnetic part of the Weyl tensor vanishes, this last result was known for perfect fluids \cite{glass}.

Now, the remaining relevant question is: what can we infer about the shear from the vanishing of  $H$ and $q$?
 We were unable  to elucidate this question in the general case, therefore in what follows, we shall restrict  to the geodesic case ($a=0$).
Under this latter condition
\begin{equation}
a=\frac{A^\prime}{A^2}=0 \quad \Rightarrow \quad  A=A(t).\label{ac}
\end{equation}

Then  assuming  $q_\alpha=H_{\alpha
\beta}=0$, and  taking into account (\ref
{ac}), we obtain from (\ref{14}) and (\ref{HH}) 
\begin{equation}
\frac{\dot B^\prime}{B}+\frac{\dot C^\prime}{C}-\frac{\dot
A}{A}\left(\frac{B^\prime}{B}+\frac{C^\prime}{C}\right)=0,\label {aq1}
\end{equation}
\begin{equation}
H=-\frac{1}{2 A^2}\left[\frac{\dot B^\prime}{B}-\frac{\dot
C^\prime}{C}+\frac{\dot
A}{A}\left(-\frac{B^\prime}{B}+\frac{C^\prime}{C}\right)\right]=0.\label{aH}
\end{equation}
Combining and integrating the equations (\ref{aq1}) and
(\ref{aH}) we obtain
\begin{eqnarray}
B(t,r)=A(t)b(r)+\alpha(t),\label{otra}
\\
C(t,r)=A(t)c(r)+\beta(t), \label{aBC}
\end{eqnarray}
where $b, c, \alpha, \beta$ are arbitrary functions of their argument.

Next, from the regularity condition $C(t,0)=0$ and redefining the function $c(r)$ we may write
\begin{equation}
C(t,r)=A(t)c(r).  \label{alCnueva}
\end{equation}

From the above equation and (\ref{sk}) it follows that $\sigma_k=0$, then (\ref{e6}) and (\ref{e7}) become
\begin{equation}
\nabla (2\Theta+\sigma _s)+3\sigma _s \frac{B^\prime}{AB}=0, \label{e6n}
\end{equation}
\begin{equation}
\nabla \sigma_s+ \sigma_s \frac{B^\prime}{AB}=0\label{e7n}.
\end{equation}
Integrating (\ref{e7n}) we obtain
\begin{equation}
\sigma _s=\frac{f(t)}{B}. \label{e6nn}
\end{equation}
where  $f(t)$ is an arbitrary integration function. However since we have $\sigma_s(t,0)=0$ we must put $f(t)=0$ implying that $\sigma_s=0$ too. Also, from (\ref{e6n}) $\nabla \Theta=0$.

Inversely, if the fluid is shearfree and geodesic then it is necessarily nondissipative.
Indeed, if the fluid is shearfree this condition can be integrated (in general, not only in the geodesic case) to obtain (see \cite{17cil} for details)
\begin{equation}
B=Ab(r) \quad {\rm and} \quad C=Ac(r).\label{bcs}
\end{equation}
where $b$ and $c$ are arbitrary functions of $r$.

Then, feeding back the above expressions for $B$ and $C$ into (\ref{8}) and taking into account (\ref{ac}) it follows that the expansion scalar is homogeneous ($\nabla \Theta=0$), implying because of (\ref{e6}) that geodesic shearfree fluid is necessarily non--dissipative.

Thus we have proved that for the geodesic fluid, $ q_\alpha=H_{\alpha
\beta}=0\Leftrightarrow \sigma_{\alpha\beta}=0$. Whereas in the general case we have stablished that $ \sigma_{\alpha\beta}=0 \Rightarrow \frac{2}{3}\nabla \Theta-\kappa q=H_{\alpha
\beta}=0$.

\section{The inhomogeneity factor and its evolution}

In the spherically symmetric case it has been shown that in the absence of dissipation the necessary and sufficient condition for the vanishing of the (invariantly defined) spatial derivative of  the energy density  is the vanishing of the scalar  associated to the trace free part of  $X_{\alpha \beta}$. For obvious reasons such a quantity was  called the inhomogeneity factor.
In other words, the inhomogeneity factor (say $\Psi$) is that combination of physical and geometric  variables, 
such  that  its vanishing is a necessary and sufficient condition for the homogeneity of energy 
density (if dissipation is present then  additional terms  including dissipative flux appear \cite{LHH}).

The extension of such a definition to situations where there is only one relevant spatial coordinate (as in the case considered here) is rather obvious:

$\Psi=0 \Leftrightarrow \nabla \mu=0$.

The two equations (\ref{e8}) and (\ref{e12}) determine the inhomogeneity factor, 
whereas (\ref{e9}) and (\ref{e10}) describes its evolution.

Nevertheless, in the present case  the situation is much more complicated (than in the spherically symmetric case) due to the fact that not only the tensor $X_{\alpha \beta}$ is expressed in terms of  three structure scalars (instead of two in the spherically symmetric case) but also due to the fact that the magnetic part of Weyl tensor is in general not vanishing and appears as a possible source of energy density inhomogeneity.

Because of the above reasons, we were unable (in the general case) to identify explicitly the inhomogeneity factor. The very simplified cases where this was possible are not very  interesting and therefore we shall not include them here.
\section{All static anisotropic cylinders}
We shall show in this section that all possible solutions of the static case are completely determined by a triplet of structure scalars. This is a reminiscence of the spherically symmetric case, where all possible static solutions are determined by a couple of structure scalars \cite{1cil}.

In the static case the field equations read:
\begin{equation}
          -\frac{B^{\prime\prime}}{B}-\frac{C^{\prime\prime}}{C}
+\frac{A^{\prime}}{A}\left(\frac{B^{\prime}}{B}+\frac{C^{\prime}}{C}\right)
   -\frac{B^{\prime}}{B}\frac{C^{\prime}}{C}=\kappa\mu A^2,
   \label{13e}
\end{equation}
\begin{equation}
\frac{A^{\prime}}{A}\left(\frac{B^{\prime}}{B}
  +\frac{C^{\prime}}{C}\right)+\frac{B^{\prime}}{B}\frac{C^{\prime}}{C}=\kappa
  P_rA^2, \label{15e}
 \end{equation}
 \begin{equation}
\frac{A^{\prime\prime}}{A}
  +\frac{C^{\prime\prime}}{C}
  -\left(\frac{A^{\prime}}{A}\right)^2  =\kappa P_zA^2, \label{16e}
  \end{equation}
\begin{equation}
\frac{A^{\prime\prime}}{A}+\frac{B^{\prime\prime}}{B}
  -\left(\frac{A^{\prime}}{A}\right)^2=\kappa P_{\phi}A^2.
  \label{17e}
\end{equation}

Then introducing the auxiliary variables
\begin{equation}
\omega=\frac{A^\prime}{A}, \qquad \xi=\frac{B^\prime}{B}, \qquad \zeta=\frac{C^\prime}{C},
\label{xyz}
\end{equation}
we can write the field equations as:
\begin{equation}
-\xi^\prime -\xi^2-\zeta^\prime -\zeta^2 + \omega\xi +\omega\zeta -\xi\zeta = \kappa \mu A^2,
\label{Ia}
\end{equation}
\begin{equation}
\omega\xi+\omega\zeta+\xi\zeta = \kappa P_r A^2,
\label{IIa}
\end{equation}
\begin{equation}
\omega^\prime +\zeta^\prime + \zeta^2 = \kappa P_z A^2,
\label{IIIa}
\end{equation}
\begin{equation}
\omega^\prime + \xi^\prime + \xi^2 =  \kappa P_\phi A^2,
\label{IVa}
\end{equation}
or,
\begin{equation}
\omega^\prime + \omega\xi +\omega\zeta  = Y_T A^2,
\label{Iaa}
\end{equation}
\begin{equation}
\omega^\prime + \xi^\prime + \xi^2 -\omega\xi -\omega\zeta -\xi\zeta =  \kappa (P_\phi - P_r) A^2,
\label{IIaa}
\end{equation}
\begin{equation}
\omega^\prime +\zeta^\prime + \zeta^2 -\omega\xi -\omega\zeta -\xi\zeta =  \kappa (P_z -P_r) A^2,
\label{IIIaa}
\end{equation}
\begin{equation}
 \xi^\prime + \xi^2 - \zeta^\prime - \zeta^2=  \kappa (P_\phi-P_z) A^2.
\label{IVaa}
\end{equation}

The scalars $E_s$ y $E_k$ take the form
\begin{eqnarray}
E_s &=& \frac{1}{2 A^2}\left(-\omega^\prime+\zeta^\prime+\zeta^2+\omega\xi-
\omega\zeta-\xi\zeta\right),\label{Essa}\\
E_k &=& \frac{1}{2 A^2}\left(-\omega^\prime+\xi^\prime+\xi^2-\omega\xi+
\omega\zeta-\xi\zeta\right).\label{Eksa}
\end{eqnarray}

Then, using (\ref{ortc2}), (\ref{YY}), (\ref{IIaa}), (\ref{IIIaa}), (\ref{Essa}) and (\ref{Eksa})
we can write
\begin{eqnarray}
Y_s A^2&=& -\omega^\prime + \omega\xi ,
\label{cpa}
\\
Y_k A^2&=& -\omega^\prime + \omega\zeta.
\label{bpa}
\end{eqnarray}
Integrating (\ref{cpa}) we obtain
\begin{equation}
A=\alpha \exp{\int{B\left(\int{\frac{-Y_s A^2}{B}dr}\right)dr}},
\label{Asol}
\end{equation}
where $\alpha$ is a constant.
Thus for any given $Y_s$,  we obtain from (\ref{Asol}) a relationship between $A$ and $B$, 
\begin{equation}
 B=B(A)\Rightarrow \omega=\omega(\xi).
 \label{BA}
 \end{equation} 
Next, from (\ref{bpa})
\begin{equation}
A=\gamma \exp{\int{C\left(\int{\frac{-Y_k A^2}{C}dr}\right)dr}},
\label{Asol1}
\end{equation}
where $\gamma$ is a constant. 
Therefore, giving $Y_k$ we obtain a relationship between $A$ and $C$,
\begin{equation}
 C=C(A)\Rightarrow \omega=\omega(\zeta).
 \label{CA}
 \end{equation} 
 Then from (\ref{BA}) and (\ref{CA}) we can express any of ($\omega, \xi, \zeta$) in terms of 
the other two.
 Therefore in (\ref{IVaa}) we can express $\xi$ and $\zeta$ in terms of $\omega$, 
obtaining a differential equation for $\omega$. This can be solved for a given $P_\phi-P_z$ and 
once $\omega$ is obtained, we can get $\xi$ and $\zeta$ from (\ref{BA}) and (\ref{CA}).

Alternatively we may use  (\ref{IIaa}) or (\ref{IIIaa})  in which case we should provide either $P_\phi - P_r$ or  $P_z -P_r$.
 Once the metric functions are found we can obtain physical variables from field equations.

Now, from (\ref{YY}) and (\ref{XX}) it follows 
\begin{equation}
\kappa(P_\phi-P_r)=-(Y_k+X_k),
\label{stn1}
\end{equation}
and  from (\ref{ortc2}) and (\ref{ortc4})
\begin{equation}
\kappa(P_z-P_r)=-(Y_s+X_s).
\label{stn2}
\end{equation}
Therefore any static anisotropic solution is determined by a triplet of scalars $(Y_k, Y_s, X_k)$ or $(Y_k, Y_s, X_s)$.

We shall next consider some special cases.

\subsection{Isotropic cylinders}
In this case  $P_r=P_z=P_\phi=P$ and field equations (\ref{15e})--(\ref{17e})  become:

\begin{eqnarray}
\frac{A^{\prime}}{A}\left(\frac{B^{\prime}}{B}
  +\frac{C^{\prime}}{C}\right)+\frac{B^{\prime}}{B}\frac{C^{\prime}}{C}&=&\kappa
  P A^2, \label{15ei}\\
  +\frac{A^{\prime\prime}}{A}
  +\frac{C^{\prime\prime}}{C}
  -\left(\frac{A^{\prime}}{A}\right)^2  &=&\kappa P A^2, \label{16ei}\\
  +\frac{A^{\prime\prime}}{A}+\frac{B^{\prime\prime}}{B}
  -\left(\frac{A^{\prime}}{A}\right)^2&=&\kappa P A^2.
  \label{17ei}
\end{eqnarray}
Then, from  (\ref{16ei}) and  (\ref{17ei}) we obtain
\begin{equation}
\frac{C^{\prime\prime}}{C}=\frac{B^{\prime\prime}}{B}, \label{BC}
\end{equation}
which in terms of the auxialiary variables introduced before, read
\begin{equation}
\zeta^\prime +\zeta^2=\xi^\prime+\xi^2,\label{yz1}
\end{equation}
which is a  Ricatti equation for  $\zeta$ ( or $\xi$).

The general solution of (\ref{yz1}) takes the form
\begin{equation}
\zeta=\xi+\frac{1}{k(r)},
\label{zy}
\end{equation}
 with
\begin{equation}
 k(r)=e^{2\int \xi dr}\left (\int
e^{-2\int \xi dr}dr+\alpha\right ).\label{yz2}
\end{equation}
Turning back to  $B$ and  $C$  we can integrate 
(\ref{zy}), to obtain
\begin{equation}
C=\beta B e^{\int \frac{1}{B^2\left(\int \frac{dr}{ B^2}+\alpha\right)}dr},\label{yz3}
\end{equation}
where $\beta$ is a constant.
From regularity conditions we must impose  that $C(t,0)=0$ for any acceptable solution.

\subsection{Conformally flat solutions}
In the static case we obtain from  (\ref{EE}) 
\begin{widetext}
\begin{equation}
E_s=\frac{1}{2A^2}\left[\frac{A^\prime}{A}\left(\frac{B^\prime}{B}+\frac{A^\prime}{A}-\frac{C^\prime}{C}\right)
-\frac{B^\prime}{B}\frac{C^\prime}{C}-\frac{A^{\prime\prime}}{A}+\frac{C^{\prime \prime}}{C}\right],\label{Es}
\end{equation}
\begin{equation}
E_k=-\frac{1}{2A^2}\left[\frac{A^\prime}{A}\left(\frac{B^\prime}{B}-\frac{A^\prime}{A}-\frac{C^\prime}{C}\right)+
\frac{B^\prime}{B}\frac{C^\prime}{C}+\frac{A^{\prime\prime}}{A}-\frac{B^{\prime \prime}}{B}\right].\label{Ek}
\end{equation}
\end{widetext}
Then, defining two functions $c(r)$ and $b(r)$ such that $C=Ac(r)$ and  $B=Ab(r) $ the conformal flatness condition ($E_s=E_k=0$) implies
\begin{equation}
c^{\prime \prime}=\frac{b^{\prime}}{b}c^{\prime},
\label{EsEk}
\end{equation}
 and 
\begin{equation}
 b^{\prime \prime}=\frac{c^{\prime}}{c}b^{\prime}. 
\label{cf1n}
\end{equation}
From (\ref{cf1n}) we get $c^{\prime}=\gamma b^{\prime}$ (where $\gamma$ is a constant) and feeding this back into (\ref{EsEk}) we obtain
\begin{equation}
b(r)=\beta \cos{(\epsilon r)},\label{cb}
\end{equation}
where $\epsilon$ and $\beta$ are constants, and the regularity condition $c(0)=0$ has been used.

In the isotropic case we obtain from (\ref{15ei}--\ref{17ei})
\begin{equation}
A= constant ;\qquad \mu=-3 P=constant.
\label{p5}
\end{equation}

All these, conformally flat, solutions have been described in detail in  \cite{cilest}, including the discussion on  junction (Darmois) conditions on the boundary surface. 
\section{Discussion and summary of results}

A comprehensive study on   cylindrically symmetric relativistic fluids 
by means of structure scalars  have been carried out. 

We have first defined the complete set of such scalars corresponding to our problem. 
It turns out that there are eight  structure scalars ($X_T, X_k, X_s, Y_T, Y_s, Y_k, Z_q, Z_H$) in contrast with the spherically symmetric  
case where there are only five. Besides, two scalars defining the shear tensor ($\sigma_k, \sigma_s$)  and three scalars defining the electric and magnetic parts of Weyl tensor ($E_k, E_s, H$) were also introduced.

Next we have stablished a set of equations governing   
the structure and evolution of   the system under consideration and brought 
out the role of structure scalars in those equations, in order to exhibit the physical  
relevance of the former.

We have first considered the dynamical equation  (\ref{e2}) derived from conservation laws.  We have next coupled the above 
mentioned equation with a transport equation derived from a causal dissipative therory. 
The resulting equation exhibits the decreasing of the effective inertial mass term due to  thermal effects.

Next, we have  brought out the relevance of structure scalars. Our main results in this respect can be summarized as follows:
\begin{itemize}
\item Three of them  $Z_q$, $Z_H$ and $X_T$,  have an evident physical meaning, and therefore do not require further discussion.
\item $Y_T$ has been shown to control the evolution of the expansion scalar through the Raychaudhuri equation (\ref{e3}), whereas $Y_s$ and $Y_k$ control the evolution of the shear through (\ref{es5}) and (\ref{es4}).
\item A very  tight link between the shearfree condition, $Z_q$ and $Z_H$ appears from equations (\ref{e6}) and (\ref{e7}). Thus, it has been shown in the geodesic case that    necessary and sufficient conditions for  the fluid to be shearfree are $Z_q=Z_H=0$.
In the general case it has been shown that  the shearfree condition implies the vanishing of the magnetic part of the Weyl tensor and a direct relationship between the inhomogeneity of the expansion scalar and the dissipative flux. The former  result explains the absence of gravitational radiation in the shearfree case already commented in \cite{17cil}.
\item The two equations (\ref{e8}) and (\ref{e12}) relates $X_s$, $X_k$, $Y_s$, $Y_k$ with energy
 density inhomogeneity and therefore should provide a definition of the inhomogeneity factor(s)
 in terms of some structure scalars. In the same order of ideas equations (\ref{e9}) and (\ref{e10})
 describe the evolution of such factor(s). Unfortunately, in the general case, we were unable
 to isolate them (it was possible only in very simplified situations).
\item We have next considered the static case. 
The main result from this section is the obtained procedure allowing to determine 
any possible solution in terms of a triplet of structure scalars $(Y_k, Y_s, X_k)$ or $(Y_k, Y_s, X_s)$. 
Particular subcases such as isotropic  or  conformally flat cylinders were studied, obtaining specific 
restrictions about  the existence of solutions satisfying regular conditions on the symmetry axis.
\end{itemize}

As expected, in a general study as the one presented here,  a great deal of questions remains  unanswered. 
Thus before ending we would like to present a partial list of issues that should be addressed in the future:
\begin{itemize}

\item From (\ref{SP}) it follows that the ``gravitational'' term vanishes not only if $H=0$ but also if $E_k=E_s$. Why?  What else does this latter condition imply?

\item We have seen that the shearfree condition implies the vanishing of the magnetic part of the Weyl tensor. What are the implications on the shear of the vanishing of the magnetic part of the Weyl tensor in the general case?

\item Are there purely magnetic solutions? 

\item Could it be possible to find the exact solution corresponding to nondissipative dust with shear (the analog of the Lemaitre--Tolman--Bondi solution)? Would this solution have a nonvanishing magnetic part of Weyl tensor?

\item We have identified the subset of equations which should determine    the inhomogeneity factor and its evolution, but we were unable to isolate such a factor in the general case. Is this possible?
\end{itemize}

\begin{acknowledgments}
LH wishes to thank Fundaci\'on Empresas Polar for financial support and Departamento   de F\'{\i}sica Te\'orica e Historia de la  Ciencia, Universidad del Pa\'{\i}s Vasco, for financial support and hospitality. ADP  acknowledges hospitality of the
Departamento   de F\'{\i}sica Te\'orica e Historia de la  Ciencia,
Universidad del Pa\'{\i}s Vasco.  JO wishes to thank Universidad de Salamanca for financial support under grant  No. FK10.
\end{acknowledgments} 
\section* {Appendix I}
The nonzero components of the Einstein equations 
\begin{equation}
G_{\alpha\beta}=\kappa T_{\alpha\beta},\label{Einstein}
\end{equation}
 are
\begin{eqnarray}
G_{00}&=&\frac{\dot{A}}{A}\left(\frac{\dot{B}}{B}+\frac{\dot{C}}{C}\right)
           +\frac{\dot{B}}{B}\frac{\dot{C}}{C}
          -\frac{B^{\prime\prime}}{B}-\frac{C^{\prime\prime}}{C}\nonumber \\
&&
+\frac{A^{\prime}}{A}\left(\frac{B^{\prime}}{B}+\frac{C^{\prime}}{C}\right)
   -\frac{B^{\prime}}{B}\frac{C^{\prime}}{C}=\kappa\mu A^2,
   \label{13}\\
G_{01}&=&-\frac{\dot{B}^{\prime}}{B}-\frac{\dot{C}^{\prime}}{C}
  +\frac{\dot{A}}{A}\left(\frac{B^{\prime}}{B}+\frac{C^{\prime}}{C}\right)\nonumber\\
&&
+\left(\frac{\dot{B}}{B}+\frac{\dot{C}}{C}\right)\frac{A^{\prime}}{A}=-\kappa
qA^2,
  \label{14}\\
G_{11}&=&-\frac{\ddot{B}}{B}-\frac{\ddot{C}}{C}
  +\frac{\dot{A}}{A}\left(\frac{\dot{B}}{B}+\frac{\dot{C}}{C}\right)
  -\frac{\dot{B}}{B}\frac{\dot{C}}{C}
  \nonumber\\
&&+\frac{A^{\prime}}{A}\left(\frac{B^{\prime}}{B}
  +\frac{C^{\prime}}{C}\right)+\frac{B^{\prime}}{B}\frac{C^{\prime}}{C}=\kappa
  P_rA^2, \label{15}\\
G_{22}&=&\left(\frac{B}{A}\right)^2\left[-\frac{\ddot{A}}{A}
  -\frac{\ddot{C}}{C}+\left(\frac{\dot{A}}{A}\right)^2
  +\frac{A^{\prime\prime}}{A}
  +\frac{C^{\prime\prime}}{C}\right.\nonumber \\
&&\phantom{\left(\frac{B}{A}\right)^2[}\left.
  -\left(\frac{A^{\prime}}{A}\right)^2\right]   =\kappa P_zB^2, \label{16}\\
G_{33}&=&\left(\frac{C}{A}\right)^2\left[-\frac{\ddot{A}}{A}
  -\frac{\ddot{B}}{B}+\left(\frac{\dot{A}}{A}\right)^2
  +\frac{A^{\prime\prime}}{A}+\frac{B^{\prime\prime}}{B}\right.\nonumber\\
&&\phantom{\left(\frac{B}{A}\right)^2[}\left.
  -\left(\frac{A^{\prime}}{A}\right)^2\right]=\kappa P_{\phi}C^2.
  \label{17}
\end{eqnarray}

\section*{Appendix II}
The non null components of the Weyl tensor
$C_{\alpha\beta\gamma\delta}$ for (\ref{3}) are
\begin{widetext}
\begin{eqnarray}
C_{0101}&=&-\frac{A^2}{6}\left[2\frac{\ddot A}{A}-\frac{\ddot
B}{B}-\frac{\ddot C}{C}-2\left(\frac{\dot A}{A}\right)^2
+2\frac{\dot B}{B}\frac{\dot C}{C} 
-2\frac{A^{\prime\prime}}{A}+\frac{B^{\prime\prime}}{B}+\frac{C^{\prime\prime}}{C}
+2\left(\frac{A^{\prime}}{A}\right)^2
-2\frac{B^{\prime}}{B}\frac{C^{\prime}}{C}\right] \nonumber \\&&=-\left(\frac{A^2}{BC}\right)^2C_{2323},
\label{1w} 
\end{eqnarray}
\end{widetext}
\begin{widetext}
\begin{eqnarray}
C_{0202}&=&\frac{B^2}{6}\left[\frac{\ddot A}{A}-\frac{2\ddot B}{B}+
\frac{\ddot C}{C}-\left(\frac{\dot A}{A}\right)^2+\frac{3\dot
A}{A}\left(\frac{\dot B}{B}-\frac{\dot C}{C}\right)
+\frac{\dot B}{B}\frac{\dot C}{C} \right. \nonumber\\
&&\left.-\frac{A^{\prime\prime}}{A}-\frac{B^{\prime\prime}}{B}+\frac{2C^{\prime\prime}}{C}
+\left(\frac{A^{\prime}}{A}\right)^2
+\frac{3A^{\prime}}{A}\left(\frac{B^{\prime}}{B}-\frac{C^{\prime}}{C}\right)
-\frac{B^{\prime}}{B}\frac{C^{\prime}}{C}\right]\nonumber \\=&-&\left(\frac{B}{C}\right)^2C_{1313},
\label{2w}
\end{eqnarray}
\end{widetext}
\begin{widetext}
\begin{eqnarray}
C_{0212}=-\frac{B^2}{2}\left[\frac{{\dot B}^{\prime}}{B}-\frac{{\dot
C}^{\prime}}{C} 
-\frac{\dot
A}{A}\left(\frac{B^{\prime}}{B}-\frac{C^{\prime}}{C}\right)-
\left(\frac{\dot B}{B}-\frac{\dot
C}{C}\right)\frac{A^{\prime}}{A}\right]=-\left(\frac{B}{C}\right)^2C_{0313},
\label{3w}
\end{eqnarray}
\end{widetext}
\begin{widetext}
\begin{eqnarray}
C_{0303}&=&\frac{C^2}{6}\left[\frac{\ddot A}{A} +\frac{\ddot
B}{B}-2\frac{\ddot C}{C} -\left(\frac{\dot
A}{A}\right)^2-3\frac{\dot A}{A}\left(\frac{\dot B}{B} -\frac{\dot
C}{C}\right)+\frac{\dot
B}{B}\frac{\dot C}{C}\right. \nonumber\\
&&\left.-\frac{A^{\prime\prime}}{A}+2\frac{B^{\prime\prime}}{B}-\frac{C^{\prime\prime}}{C}+
\left(\frac{A^{\prime}}{A}\right)^2-
3\frac{A^{\prime}}{A}\left(\frac{B^{\prime}}{B}-\frac{C^{\prime}}{C}\right)-
\frac{B^{\prime}}{B}\frac{C^{\prime}}{C}\right]\nonumber \\&=&-\left(\frac{C}{B}\right)^2C_{1212}.
\label{4w}
\end{eqnarray}
\end{widetext}
The  two components of the Riemann tensor appearing in (\ref{UastR}) are
\begin{equation}
R_{0202}=-B^2\left(\frac{\ddot B}{B}-\frac{\dot A}{A}\frac{\dot
B}{B}-\frac{A^{\prime}}{A}\frac{B^{\prime}}{B}\right), \\ \label{8w}
\end{equation}
\begin{equation}
R_{2323}=\left(\frac{BC}{A}\right)^2\left(\frac{\dot B}{B}\frac{\dot
C}{C}-\frac{B^{\prime}}{B}\frac{C^{\prime}}{C}\right). \label{13w}
\end{equation}
\section{Appendix III}
From their definition it is not difficult to express structure scalars through metric functions and their derivatives, these expressionas are:
\begin{widetext}
\begin{equation}
X_T=\frac{1}{A^2}\left(\frac{\dot A \dot B}{AB}+\frac{\dot A \dot C}{AC}
+\frac{\dot B \dot C}{BC}-\frac{B^{\prime \prime}}{B}-\frac{C^{\prime \prime}}{C}
+\frac{A^\prime B^\prime}{AB} +\frac{A^\prime C^\prime}{AC}-\frac{B^\prime C^\prime}{BC}\right),
\label{XTm}
\end{equation}
\end{widetext}
\begin{equation}
X_s=\frac{1}{A^2}\left(\frac{\dot A \dot C}{AC}-\frac{\dot B \dot C}{BC}-\frac{C^{\prime \prime}}{C} 
+\frac{A^\prime C^\prime}{AC}+\frac{B^\prime C^\prime}{BC}\right),
\label{Xsm}
\end{equation}
\begin{equation}
X_k=\frac{1}{A^2}\left(\frac{\dot A \dot B}{AB}-\frac{\dot B \dot C}{BC}-\frac{B^{\prime \prime}}{B} 
+\frac{A^\prime B^\prime}{AB}+\frac{B^\prime C^\prime}{BC}\right),
\label{Xkm}
\end{equation}
\begin{widetext}
\begin{equation}
Y_T=\frac{1}{A^2}\left(-\frac{\ddot A}{A}-\frac{\ddot B}{B}-\frac{\ddot C}{C}+\frac{\dot A^2}{A^2}
+\frac{\dot A \dot B}{AB}+\frac{\dot A \dot C}{AC}+\frac{A^{\prime \prime}}{A}-\frac{{A^\prime}^2}{A^2}
+\frac{A^\prime B^\prime}{AB} +\frac{A^\prime C^\prime}{AC}\right),
\label{YTm}
\end{equation}
\end{widetext}
\begin{widetext}
\begin{equation}
Y_s=\frac{1}{A^2}\left(\frac{\ddot A}{A}-\frac{\ddot B}{B}-\frac{\dot A^2}{A^2}+\frac{\dot A \dot B}{AB}
-\frac{A^{\prime \prime}}{A}+\frac{{A^\prime}^2}{A^2}+\frac{A^\prime B^\prime}{AB}\right),
\label{Ysm}
\end{equation}
\end{widetext}
\begin{widetext}
\begin{equation}
Y_k=\frac{1}{A^2}\left(\frac{\ddot A}{A}-\frac{\ddot C}{C}-\frac{\dot A^2}{A^2}+\frac{\dot A \dot C}{AC}
-\frac{A^{\prime \prime}}{A}+\frac{{A^\prime}^2}{A^2}+\frac{A^\prime C^\prime}{AC}\right).
\label{Ykm}
\end{equation}
\end{widetext}

\end{document}